\documentclass[twocolumn,aps,prd,longbibliography, nobibnotes,byrevtex,nofootinbib,superscriptaddress,amsfonts,amsmath,amssymb,floatfix,10pt]{revtex4-2}
\UseRawInputEncoding
\usepackage{appendix}
\usepackage{hyperref}
\usepackage{enumerate}
\hypersetup{
    colorlinks=true,
    citecolor=blue,
    linkcolor=blue,
    filecolor=magenta,      
    urlcolor=blue,
}
\usepackage{times}
\hypersetup{unicode=true}
\usepackage{graphicx}
\usepackage{dcolumn}
\usepackage{mathrsfs}
\usepackage{acronym}
\usepackage[usenames,dvipsnames,svgnames]{xcolor}
\usepackage{tabularx}
\usepackage{subfigure}

\newcommand{\msun}{\ensuremath{\mathrm{M}_{\odot}}}

\begin{document}
\title{Vanishing Compactness Gap and Fermionic Compact Dark Matter in Ho\v{r}ava-Lifshitz Gravity}

\author{Edwin J. Son}
\email[]{eddy@nims.re.kr}
\affiliation{National Institute for Mathematical Sciences, Daejeon 34047, Republic of Korea}

\author{Kyungmin Kim}
\email[]{kkim@kasi.re.kr}
\affiliation{Korea Astronomy and Space Science Institute, Daejeon 34055, Republic of Korea}

\author{John J. Oh}
\email[]{johnoh@nims.re.kr}
\affiliation{National Institute for Mathematical Sciences, Daejeon 34047, Republic of Korea}

\date{\today}

\begin{abstract}
We show that the gap in the compactness between black holes and neutron stars witnessed in general relativity may be vanishing in \ac{hl} gravity. Assuming a fermion equation-of-state for simplicity, and solving the \acl*{tov} equation within the \ac{hl} gravity framework, we see that there exists a minimum fermion mass $m_f^\text{(min)}(q,y)$, above which the gap of the compactness between black hole and fermionic compact object vanishes, for a given deformation parameter $q$ of \ac{hl} and interaction strength $y$ between fermions. Thus, in \ac{hl} gravity, the mass and radius of an object found in the lower mass gap by LIGO-Virgo-KAGRA observations might not be able to classify it as a black hole or a neutron star. It is interesting to note that a fermion of mass $\sim 40\ \text{GeV}$ can form a highly compact object of mass $\sim 10^{-4}\ \msun$ and radius $\sim 1\ \text{m}$ that may play the role of the cold dark matter. In addition, we find the possible existence of another class of compact objects whose compactness is comparable to that of a black hole.
\end{abstract}

\maketitle

\acrodef{hl}[HL]{Ho\v{r}ava-Lifshitz}
\acrodef{ks}[KS]{Kehagias-Sfetsos}
\acrodef{tov}[TOV]{Tolman-Oppenheimer-Volkoff}
\acrodef{gr}[GR]{general relativity}
\acrodef{uv}[UV]{ultraviolet}
\acrodef{ir}[IR]{infrared}
\acrodef{EOS}[EOS]{equation-of-state}
\acrodef{ns}[NS]{neutron star}
\acrodef{bh}[BH]{black hole}
\acrodef{apr4}[APR4]{Akmal-Pandharipande-Ravenhall}
\acrodef{dm}[DM]{dark matter}
\acresetall

\section{Introduction}
\label{sec:intro}

\ac{hl} gravity~\cite{Horava:2008ih,Horava:2009uw,Horava:2009if} has been proposed as a framework in which gravitational dynamics can vary with the energy scale. 
While \ac{gr} successfully describes gravity at macroscopic scales, it fails to remain renormalizable in the \ac{uv} regime, where quantum effects become unavoidable—for example, in the early Universe or near \ac{bh} horizon. 
A renormalizable theory of gravity is therefore expected to exhibit improved behavior at high energies and short distances.
\ac{hl} gravity approaches this challenge by abandoning local Lorentz invariance at high energies and introducing an anisotropic scaling between time and space, reminiscent of Lifshitz-type theories in condensed-matter physics. This scaling modification is crucial for achieving power-counting renormalizability, even though it explicitly breaks Lorentz symmetry in the \ac{uv} regime. Importantly, \ac{hl} gravity is constructed so that GR is recovered in the infrared (IR) limit, ensuring consistency with well-tested low-energy gravitational phenomena.
Since its inception, \ac{hl} gravity has inspired extensive research across cosmology \cite{Mukohyama:2010xz}, \ac{bh} physics \cite{Kehagias:2009is, PhysRevLett.103.091301}, and quantum gravity as a candidate \ac{uv}-complete theory of gravitation \cite{Wang:2017brl}.

In particular, the deformed \ac{hl} gravity can be treated as a special model since it possesses an asymptotically Minkowskian structure while exhibiting modifications in the \ac{uv} regime~\cite{Horava:2009uw,Park:2009zra}. In Ref.~\cite{Kimetal2021}, the authors investigated the static configurations of \ac{ns}-like objects with several representative \ac{EOS} in the deformed \ac{hl} gravity, finding that the maximum stable mass and radius become larger than those in \ac{gr} due to the effectively weaker gravitational interaction in the \ac{uv} region. This naturally raises a question: \emph{Can purely fermionic matter form compact objects with \ac{ns}-scale masses in the deformed \ac{hl} framework?} 

Meanwhile, in \ac{gr}, there exists a gap in the allowed values of the compactness $M / R$ with $c=G=1$, where $M$ and $R$ denote the mass and radius of a self-gravitating system, respectively. The compactness of a Schwarzschild \ac{bh} is always 0.5, while that of a typical \ac{ns} is predicted as about 0.1--0.2~\cite{Oppenheimer:1939ne,Rhoades:1974fn,Kalogera:1996ci}, regarding possible channels for producing each type of compact objects. The theoretical maximum of a \ac{ns} is about 0.3~\cite{Rhoades:1974fn,Kalogera:1996ci} and no compact object can be formed with compactness between 0.3 and 0.5 in \ac{gr}.\footnote{Though the upper bound in the compactness of charged compact objects for a Reissner-Nordstr\"om vacuum is 8/9~\cite{Dadhich:2019jyf,Giuliani:2007zza}, the observed \acp{ns} are charge neutral so that such charged compact objects are not under consideration in this work. In addition, we only consider non-rotating objects for simplicity.}
This naturally provokes another question: {\it Is the compactness gap itself a universal prediction of gravity models, or is it a theory-dependent effective phenomenon?}

In relation to the compactness gap defined theoretically, there exists observational evidence of the so-called \emph{lower mass gap} between \acp{ns} and \acp{bh}, typically in the range of $3\ \msun$ to $5\ \msun$~\cite{Ozel_2010,Fishbach_2020,Fishbach_2024}.\footnote{Note that the lower ends of both the compactness gap and the mass gap correspond to the maximum mass (or maximum compactness) of \acp{ns} in \ac{gr}}.
As an example, the primary mass of the compact binary system associated with the gravitational-wave event, GW230529\_181500, has been inferred to be $2.5$--$4.5~\msun$~\cite{LIGOScientific:2024elc}.
Relevant studies~\cite{Fishbach2017, GerosaBerti2017, Margalit:2017dij, Abbott2020Population, Fishbach2020, Fishbach2024} discuss the possibility that objects within the observed mass gap may result from binary merger processes, rather than originating from isolated stellar evolution. 
However, the merger rate of \iac{ns} and the remnant of a binary neutron star merger is several orders of magnitude lower than the rate inferred for the event like GW230529\_181500~\cite{LIGOScientific:2024elc}. 
Therefore, a debate about the unresolved lower mass gap in astrophysical observations---closely related to the compactness gap in \ac{gr}---is still ongoing.

Motivated by this, we demonstrate that the compactness gap between \acp{ns} and \acp{bh} can vanish in the framework of deformed \ac{hl} gravity. By solving the \ac{tov} equations~\cite{Tolman:1939jz,Oppenheimer:1939ne} for fermionic matter within the framework of deformed \ac{hl} gravity, we show that there exist stable compact configurations whose mass and radius approach those of the minimal \ac{ks} \ac{bh}~\cite{Kehagias:2009is}, thereby continuously filling the gap in compactness. This result suggests that, once gravity is modified, the notion of a compactness gap may not be guaranteed, and that compact objects with compactness comparable to \acp{bh} can be supported without forming an event horizon. As a consequence, compact objects lying in the compactness gap could be interpreted as exotic compact objects arising from modified gravity. 

Furthermore, for certain ranges of fermion masses and interaction strengths, such highly compact configurations may also appear at sub-solar masses, which suggests that they could constitute a viable class of compact \ac{dm} candidates. 
Specifically, our results 
indicate that the vanishing compactness gap represents a distinctive strong-field astrophysical signature of \ac{hl} gravity beyond GR, with potential implications for both gravitational-wave observations and \ac{dm} phenomenology.
Therefore, comparing fermionic compact \ac{dm} in \ac{gr}~\cite{Narain:2006kx, Valdez-Alvarado:2013, Mukhopadhyay:2016, Mariani:2023} with its counterpart in the deformed \ac{hl} gravity may shed light on the viability of modified gravity as a \ac{dm} candidate.

We organize this paper as follows: In Sec.~\ref{sec:HL}, we briefly introduce the deformed \ac{hl} gravity model and the corresponding \ac{tov} equation and summarize the equation of state model that we are considering. In Sec.~\ref{sec:NS}, we present the result of numerical study performed for the compactness gap and the fermionic cold \ac{dm} candidates. In Sec.~\ref{sec:discussion}, we discuss the results and present the future prospects. 

\section{TOV equation in HL gravity}
\label{sec:HL}
\subsection{Equations of motion in HL gravity}
The action of \ac{hl} gravity is formulated by an anisotropic scaling between time and space, $t \rightarrow b^{z}  t$ and $x^i \rightarrow b x^i$ and the form of $z=3$:  
\begin{eqnarray}
I_\text{\acs{hl}} &=& \int dt d^3x \sqrt{g} N \left[\frac{2}{\kappa^2} \left( {\mathcal K}_{ij}{\mathcal K}^{ij} - \lambda {\mathcal K}^2 \right) \right.\nonumber\\ 
&&-\left. \frac{\kappa^2}{2\zeta^4} \left( {\mathcal C}_{ij} - \frac{\mu\zeta^2}{2} {\mathcal R}_{ij} \right) \left( {\mathcal C}^{ij} - \frac{\mu\zeta^2}{2} {\mathcal R}^{ij} \right)\right.\nonumber\\
&&+ \! \left. \! \frac{\kappa^2\mu^2(4\lambda \! - \! 1)}{32(3\lambda \! - \! 1)} \! \! \left( \! {\mathcal R}^2 \! + \! \frac{2}{(4\lambda \! - \! 1) q^2} {\mathcal R} \! + \! \frac{12 \Lambda_W^2}{4\lambda \! - \! 1} \! \right) \! \right]
\label{action}
\end{eqnarray}
with the \emph{softly} broken detailed balance condition,
which is sometimes called the deformed \ac{hl} gravity in the literature~\cite{Horava:2008ih,Horava:2009uw,Horava:2009if,Park:2009zra}. Note that ${\mathcal K}_{ij} \equiv \frac{1}{2N} [ \dot{g}_{ij} - \nabla_i N_j - \nabla_j N_i ]$ is an extrinsic curvature, where $N$ is a lapse function, $N^{i}$ is a shift vector, $g_{ij}$ is a three-dimensional spatial metric, and $\dot{g}_{ij}$ denotes $\partial{}g_{ij}/\partial{}t$. ${\mathcal C}^{ij} \equiv \varepsilon^{ik\ell} \nabla_k ( {\mathcal R}_\ell^j -  \delta_\ell^j {\mathcal R}/4 )$ is a Cotton-York tensor, where ${\mathcal R}_{ij}$ and ${\mathcal R}$ are a three-dimensional spatial Ricci tensor and a Ricci scalar, respectively. $\kappa^2$ is a coupling constant related to the Newton's gravitational constant $G_N$, and $\lambda$ is an additional dimensionless coupling constant characterizing deviations from \ac{gr} in the kinetic term of \ac{hl} gravity.
Here, $q$ is essential for an asymptotically flat solution, though it violates the detailed balance condition. The coupling constants $\mu$, $\Lambda_W$, and $\zeta$ stem from the three-dimensional Euclidean topologically massive gravity action~\cite{Deser:1981wh, Deser:1982vy}. When $\lambda=1$, the Einstein-Hilbert action can be recovered in the \ac{ir} limit by identifying the fundamental constants with $c = (\kappa^2 / 4) [\mu^2 / 2 q^2 (3\lambda-1)]^{1/2}$, $G_N = \kappa^2 c^2 / 32 \pi$, and $\Lambda = -3 q^2 \Lambda_W^2$, 
representing the speed of light, Newton's gravitational constant, and the cosmological constant, respectively.

The total action under consideration is given by $I_\text{tot} = I_\text{\acs{hl}} + I_\text{mat}$, where $I_\text{mat}$ represents the matter action that will be specified by assuming a perfect fluid and choosing an \ac{EOS} without an explicit form. Hereafter, we consider $c=G_N=1$ and an asymptotically flat geometry, that is, $\Lambda = 0$, for simplicity.

Now, if we consider a static, spherically symmetric metric ansatz,
\begin{equation}
\label{eq:metric}
ds^2 = - e^{2\Phi(r)} dt^2 + \frac{dr^2}{f(r)} + r^2 \left( d\theta^2 + \sin^2\theta d\phi^2 \right),
\end{equation}
then the equations of motion in \ac{hl} gravity with the stress-energy tensor of a perfect fluid, $T_{\mu\nu} = (\rho+p) u_{\mu}u_{\nu} + pg_{\mu\nu}$, are given by
\begin{subequations}
\label{eom}
\begin{align}
\rho &= \frac{q^2}{8\pi r^2} \left( \frac{r}{q^2} (1 - f) + \frac{(1 - f)^2}{r} \right)', \label{eom:rho} \\
p &= \frac{q^2}{8\pi r^4} \bigg[ (1 - f) \left( 1 - f - \frac{r^2}{q^2} \right) \notag \\
  & \qquad \qquad \ + 4 r f \left( 1 - f + \frac{r^2}{2q^2} \right) \Phi' \bigg],
  \label{eom:p} \\
p' &= - \left( \rho + p \right) \Phi', \label{eom:cons}
\end{align}
\end{subequations}
where $u_{\mu}=(1,0,0,0)$ is a four-vector field, $\rho$ and $p$ are the energy density and the pressure of a perfect fluid, respectively, and the prime denotes $d/dr$. Note that \ac{hl} gravity has six parameters, as observed in Eq.~\eqref{action}: we have already fixed $\lambda$ and $\Lambda_W$ to 1 and 0, respectively, $\kappa$ and $\mu$ are hidden in the physical constants related to $c$ and $G_N$ in the \ac{ir} region, and $\zeta$ does not contribute to this non-rotating configuration. Thus, $q$ is the only remaining free parameter of the theory.

Following the procedure in Ref.~\cite{Sonlimits} to solve Eq.~\eqref{eom}, we replace $f(r)$ by $m(r)$ through the relation
\begin{equation}
  \label{eq:f}
  f = \frac{1 - 2 m/r + q^2/r^2}{2^{-1} \left[ 1 + 2q^2/r^2 + \sqrt{1 + 8q^2 m r^{-3}} \right]}.
\end{equation}
Note that the numerator is the same form of the Reissner-Nordstr\"om vacuum, when $m(r) = M$ and $q = Q$, which reminds us that the noncommutative Schwarzschild \ac{bh} behaves like the Reissner-Nordstr\"om \ac{bh} in thermodynamic analysis~\cite{Kim:2008vi}. 
Similar to the Reissner-Nordstr\"om vacuum, the horizons of the \ac{ks} \ac{bh} are located at $r_\pm =  M \pm \sqrt{M^2 - M_c^2}$, where $M_c\equiv |q|$ is the critical mass for horizon formation~\cite{Kehagias:2009is}.

However, in the limit of $q/r \ll 1$, the metric function~\eqref{eq:f} approximates $f \approx 1 - 2 M/r + 4 q^2 M^2 / r^4 + \mathcal{O}(q/r)^4$ so that it converges to the Schwarzschild vacuum either in the \ac{gr} limit ($q\to0$) or in \ac{ir} limit ($r\to\infty$) where the \ac{hl} correction terms are negligible.
Here, the mass parameter $M$ is identified as the quasilocal energy at infinity~\cite{Myung_2010}. For the \ac{ks} solution, we see $e^{2\Phi(r)} = f(r)$, while Eq.~\eqref{eom:cons} governs the behavior of the  $(tt)$-component of the metric, which is coupled to matter in the generic case. 

With the Planck mass $m_P$ and length $l_P$, $M_c$ can be rewritten as $M_c = |q| m_P / \ell_P$, which reduces to $M_c = m_P$ for $|q| = \ell_P$ and becomes $M_c \approx \msun$ for $|q| \approx 1.477 \times 10^3 \, \text{m}$, where $\msun$ represents the solar mass. For a constant $q$, the \ac{bh} horizon exists with $M\ge M_{c}$.\footnote{When $M < M_{c}$, a naked singularity appears, and the detailed behavior of the formation of naked singularities and wormholes in Ho\v{r}ava gravity is studied in~\cite{Bellorin:2014qca,Son:2010bs}. In the quantum regime, however, the \ac{ks} geometry turns out to be regular~\cite{Gurtug:2017kqf}, which may safely avoid the naked singularity problem.} On the other hand, for a fixed $M$, one finds  $|q| \le q_{c} \equiv M$ 
to form a \ac{bh}. Hence, the horizon of a solar mass \ac{ks} \ac{bh} exists only if $|q| \le q_{c}\approx 1.477 \times 10^3 \, \textrm{m}$. 
Considering the possibility of the existence of $\gtrsim 3\ \msun$ \ac{bh}, we assume $|q| \lesssim 4.5 \times 10^3 \ \textrm{m}$.
Accordingly, in this study we restrict the range of the parameter as $|q| \le 4 \times 10^3 \ \textrm{m}$.

\subsection{TOV equations for fermionic gas}
\label{sec:eos}

Now, we can rewrite Eq.~\eqref{eom} as follows:
\begin{subequations}
\label{eq_tov}
\begin{align}
m' &= 4 \pi r^2 \rho, \label{Eq_tov_mass} \\
p' &= - \frac{m \rho \mathfrak{A}}{r^2 \mathfrak{B}} \frac{(1 + p/\rho) \left[ 1+ 4 \pi r^3 p \mathfrak{B} / m - q^2 \tilde{\rho}  \right]}{\sqrt{1 + 8q^2 \tilde{\rho}} \left[ 1 - 2 m / r + q^2/r^2 \right]}, \label{eq_tov_pre}
\end{align}
\end{subequations}
where $\mathfrak{A} = 2^{-1} \big[ 1 + 2q^2/r^2 + \sqrt{1 + 8q^2 \tilde{\rho}} \big]$, $\mathfrak{B} = 2^{-1} \big[ 1 + 2q^2 \tilde{\rho} + \sqrt{1 + 8q^2 \tilde{\rho}} \big]$, and  $\tilde{\rho} = m r^{-3}$.
Note that if we expand $p'$ in terms of $q \to 0$ and take the leading order, we obtain
\begin{equation}
 p' \approx - \frac{m \rho}{r^2} \left(1 + \frac{p}{\rho}\right) \left( 1 + \frac{4 \pi r^3 p}{m} \right) \left(1 - \frac{2 m}{r}\right)^{-1},
\end{equation}
which reproduces the \ac{tov} equation in \ac{gr}.

\begin{figure*}
    \centering
    \subfigure[GR ($q=0$)]{
    \includegraphics[width=0.31\linewidth]{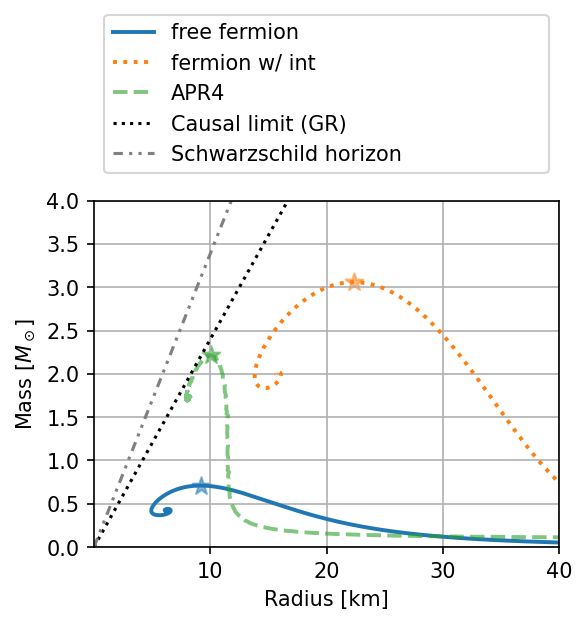}
    }
    \subfigure[HL ($q=400~\textrm{m}$)]{
    \includegraphics[width=0.31\linewidth]{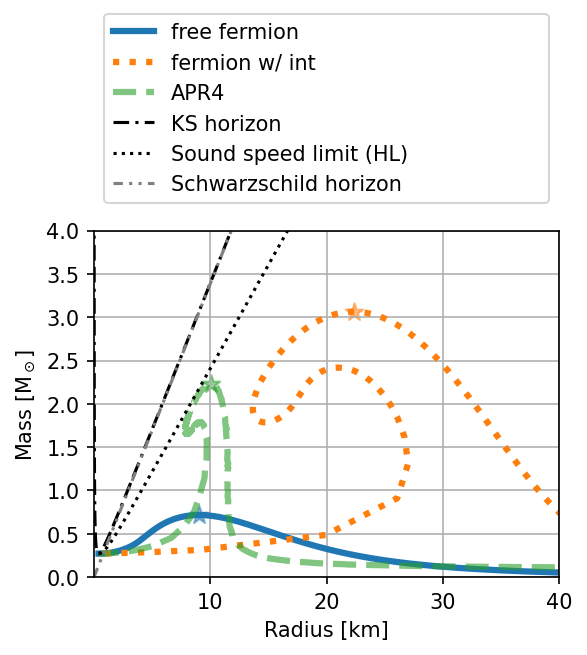}
    }
    \subfigure[HL ($q=4\,000~\textrm{m}$)]{
    \includegraphics[width=0.31\linewidth]{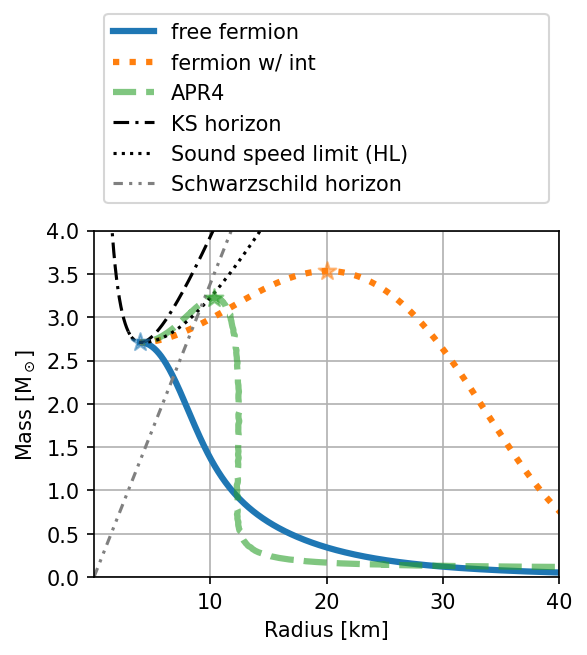}
    }
    \caption{Comparison of fermionic compact objects of $m_f = m_n$ with and without interaction term and compact object whose \acs{EOS} is \acs{apr4} in (a) \acs{gr} and (b,c) \acs{hl}. The blue solid curves represent compact objects made of free fermions, the orange dotted curves represent fermionic compact objects with two-body interactions, and the green dashed curves are the compact objects with the \acs{apr4} \acs{EOS}. The horizons of the \acs{ks} and Schwarzschild \acp{bh} are also plotted as the dash-dotted and dash-double dotted curves, respectively. The sound speed limit (causal limit) of \acs{ks} and Schwarzschild vacua are depicted as dotted curves, respectively.}
    \label{fig:mr_mn_ap4}
\end{figure*}

The \ac{EOS} for a free fermion gas at zero temperature is given by the energy density $\rho$ and the pressure $p$, defined as
\begin{eqnarray}
\rho&=& \frac{1}{\pi^2} \int_{0}^{k_{F}} k^2 \sqrt{m_{f}^2+k^2} dk\nonumber \\
&=& \frac{m_{f}^4}{8\pi^2} [(2\eta^3 + \eta)(1+\eta^2)^{1/2}-\sinh^{-1}(\eta)], \label{eq:freefermion_ene}\\
p&=& \frac{1}{3\pi^2} \int_{0}^{k_{F}} \frac{k^4}{\sqrt{m_{f}^2 + k^2}} dk \nonumber \\
&=& \frac{m_{f}^4}{24\pi^2} [(2\eta^3 - 3\eta)(1+\eta^2)^{1/2} +3\sinh^{-1}(\eta)], \label{eq:freefermion_pre}
\end{eqnarray}
where $k_f$ is the Fermi momentum, $m_f$ is the mass of a fermion, and $\eta$ is the dimensionless Fermi momentum defined as $\eta \equiv k_{F}/m_{f}$.

In a more realistic treatment, it is natural to introduce interactions between fermions. To this end, we follow the approach of \cite{Narain:2006kx}, in which two-body interaction terms are added to both Eqs.~\eqref{eq:freefermion_ene} and~\eqref{eq:freefermion_pre} as
\begin{equation}
\rho_\mathrm{int}= \left(\frac{m_f^2}{3\pi^2}\right)^2 y^2 \eta^6, ~~p_\mathrm{int} = \left(\frac{m_f^2}{3\pi^2}\right)^2 y^2 \eta^6,
\end{equation}
where $y=m_f/m_I$ parametrizes the interaction strength, with $m_I$ denoting the interaction energy scale.

\section{Numerical Solutions for Fermionic Compact Objests in HL Gravity}
\label{sec:NS}

\subsection{Fermion of neutron mass}

To obtain the mass-radius relation of fermionic compact objects in HL gravity (Fig.~\ref{fig:mr_mn_ap4}), we numerically solve the TOV equation in Eq.~\eqref{eq_tov} while varying the central energy density, $\rho_c$, for three selected values of the parameter $q$: $0~\mathrm{m}$, $400~\mathrm{m}$, and $4\,000~\mathrm{m}$.\footnote{In solving the TOV equation, as done in~\cite{Sonlimits}, we employ the eighth-order Runge-Kutta method~\cite{Hairer1993} implemented in \texttt{SciPy}~\cite{2020SciPy-NMeth}} In addition, the sound speed limit for \ac{ks} vacuum in the figure is obtained by restricting the sound speed inside the object less than or equal to the speed of light $c$, which is referred to as the causal limit in \ac{gr}~\cite{Sonlimits}. We find that all the stable compact objects are well formed below the sound speed limit curve.

As shown in the case (a) of Fig.~\ref{fig:mr_mn_ap4}, solving the TOV equation in GR---the TOV equation in HL gravity with $q=0$, equivalently---shows that a compact object supported by a free fermionic gas of neutron-like fermions ($\sim 1~\textrm{GeV}$) has a maximum mass below one solar mass, whereas the inclusion of interactions among the fermions increases both the maximum mass and the radius.~\cite{Oppenheimer:1939ne,Narain:2006kx} Here, we take $y=10$ as an example. For comparison, the mass-radius profile for a \ac{ns} \ac{EOS}, \ac{apr4}~\cite{APR:1998}, is also presented. 

In \ac{hl} gravity, i.e., for $q \neq 0$, considering the same fermion mass as that of neutron, the resulting mass-radius relations of fermionic compact objects are shown in the cases (b) and (c) of Fig.~\ref{fig:mr_mn_ap4}:
For the case of (b) $q=400~\textrm{m}$, the behavior of compact objects is almost the same as that of \ac{gr}. The main difference is exhibited on the left branch beyond the maximum point (the most massive compact object), which is unstable and thus unphysical as an equilibrium solution. 
However, for the case of (c) $q=4\,000~\textrm{m}$, the maximum mass is significantly larger than that in \ac{gr}. Furthermore, the mass and the size of the most massive compact object of free fermion are almost the same as those of the minimal \ac{ks} \ac{bh}. It is plausible that \ac{hl} gravity allows such objects which may not be distinguishable from \acp{bh} with current observational capabilities. We will look into it in the following section.

\subsection{Arbitrary \texorpdfstring{$q$}{q} and various \texorpdfstring{$m_f$}{mf}}

\begin{figure}
    \centering
    \includegraphics[width=\columnwidth]{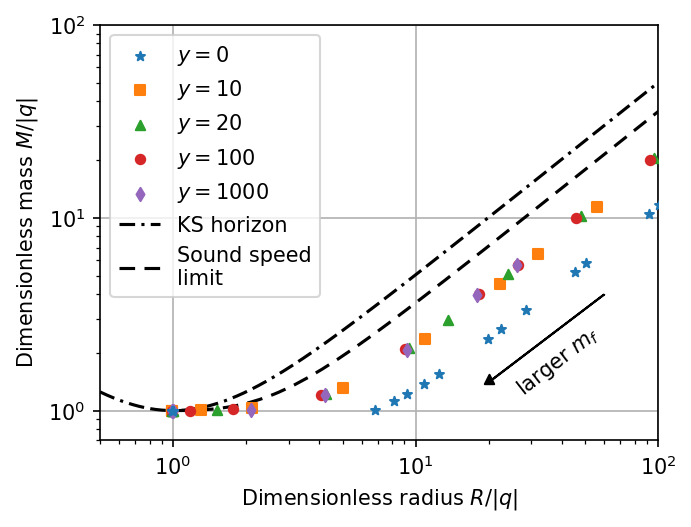}
    \makeatletter\long\def\@ifdim#1#2#3{#2}\makeatother
    \caption{Comparison of fermionic compact objects with and without two-body interaction term in \ac{EOS} for some interaction strengths $y$ and various fermion masses $m_f$. For a given $y$, the point of smaller $M/|q|$ and $R/|q|$ indicates the case of more massive $m_f$. All the most massive stable objects are below the sound speed limit (causal limit) in \acs{ks} vacuum.}
    \label{fig:mrq}
\end{figure}

We now consider an arbitrarily nonzero $q$, keeping the above constraints, of course, and introduce dimensionless mass $M/|q|$ and radius $R/|q|$. Then, the most massive stable compact objects of $q$-independent mass-radius profiles can be plotted in Fig.~\ref{fig:mrq}: Each point represents the most massive compact object when a fermion mass $m_f$ and an interaction strength $y$ are given, which are distinguished by the position and the marker shape, respectively. Specifically, for a given $y$, the point of smaller values in both $M/|q|$ and $R/|q|$ indicates the case of more massive $m_f$. From this figure, we find that all the most massive objects are formed below the sound speed limit curve~\cite{Sonlimits}---obtained in analogy with the calculation of the causal limit in \ac{gr}~\cite{Rhoades:1974fn,Kalogera:1996ci}. Then, it is implied that the causality remains satisfied for compact objects in \ac{hl}.

It is interesting to note that $R$ goes to $|q|$, as $m_f$ becomes larger, and the most massive objects of all considered values of the interaction strength $y$ come closer to the \ac{ks} horizon and eventually reach the minimal \ac{bh} point. In other words, for an arbitrary $q$, there may exist a compact object whose mass and radius are similar to the minimal \ac{bh}, $M \sim |q| \sim R$. However, we find no significant difference among the most massive objects of different nonzero $y$'s except for the case $y=0$. From this result, we understand that the $m_f$-dependent profile mainly relies on whether $y=0$ or not.

\begin{figure*}
    \centering
    \subfigure[ Compactness vs. fermion mass]{
      \includegraphics[width=\textwidth]{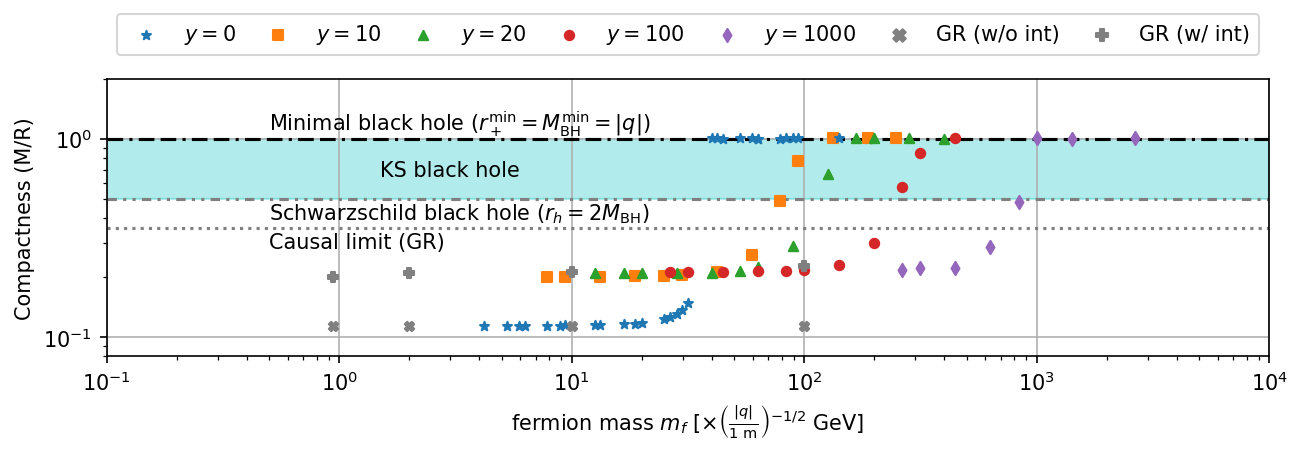}
      \label{fig:mrq_compactness_mf}
    }
    \subfigure[ Compactness vs. central density]{
      \includegraphics[width=\textwidth]{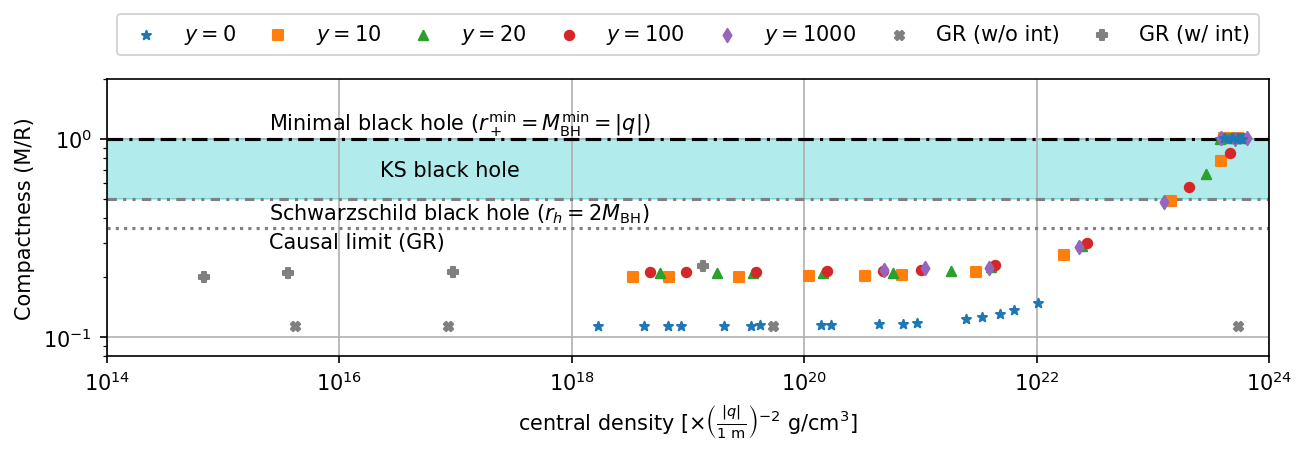}
      \label{fig:mrq_compactness_rhoc}
    }
    \caption{Compactness of the most massive fermionic compact objects in \acs{hl} gravity for several interaction strengths $y$ are depicted with respect to (a) various fermion masses $m_f$ and (b) various central densities $\rho_c$. Compactness of several fermionic compact objects in \acs{gr} with interaction energy $m_I \sim 100\ \text{MeV}$ are also shown in dimension (a) [GeV] and (b) [g/cm$^3$] for comparison.
    The compactness of a \acs{ks} \ac{bh} is between that of a Schwarzschild \ac{bh}, 1/2, and that of a minimal \ac{bh}, 1, according to the ratio of $|q|/M_\mathrm{BH}$, and it is represented by the shaded region. The compactness of sound speed limit of \ac{ks} vacuum is also between that of causal limit of Schwarzschild vacuum in \acs{gr} and that of a minimal \ac{bh}, according to the ratio of $|q|/R$.}
    \label{fig:mrq_compactness}
\end{figure*}

To see the properties of such objects, we investigate the compactness $M/R$ with respect to the fermion mass $m_f$ and the central density $\rho_c$ (Fig.~\ref{fig:mrq_compactness}). The compactness of the \ac{ks} \ac{bh} is in the range of $1/2 < M/R \le 1$: The large \ac{ks} \acp{bh} ($r_+ \gg |q|$) is approximated to Schwarzschild \acp{bh} and the small \ac{ks} \acp{bh} is bounded by the minimal \ac{bh} $r_+^\text{min} = M_\text{BH}^\text{min} = |q|$. As mentioned above, fermionic compact objects can be formed with the masses and radii similar to those of the minimal \ac{bh}, and hence their compactness is bounded from above, $M/R < 1$.

As shown in Fig.~\ref{fig:mrq_compactness_mf}, the fermion mass scale for a compact object to attain a compactness within the \ac{ks} \acp{bh} range depends on the interaction strength $y$, as summarized in Table~\ref{tab:compactness}. 
The scales in the table look somewhat high, but higher $q$ reduces the values to reasonable scales. For example, considering $q \sim 4\,000\ \text{m}$, the mass scales in Table~\ref{tab:compactness} are multiplied by a factor $\sim 1/60$ and reduce to $\sim 0.6$--16 GeV.

In Fig.~\ref{fig:mrq_compactness_rhoc}, we plot the compactness $M/R$ of the most massive fermionic compact objects in \ac{hl} gravity as a function of the central density $\rho_{c}$ for several interaction strength $y$. For a given $y$, the compactness increases monotonically with increasing central density, reflecting the formation of progressively denser equilibrium configurations.
It is interesting to note that the central density of a fermionic compact object with compactness $1/2 \le M/R \le 1$ is given by $\sim 10^{23}\times (|q|/1\ \textrm{m})^{-2}\ \textrm{g/cm}^{3}$ for all $y$, which reduces to $\sim 10^{16}\ \textrm{g/cm}^{3}$ for $q \sim 4\,000\ \textrm{m}$. Note that this critical central density is universal in the sense that it is independent of the interaction strength $y$.

\begin{table}[tbp]
    \caption{The mass scale of the fermion for a compact object to have its compactness in the range of that of \ac{ks} \acp{bh} depends on the interaction strength $y$.}
    \label{tab:compactness}
    \begin{tabular*}{\linewidth}{@{\extracolsep{\fill}} r | r @{}}
        \hline
        \hline
        interation strength $y$ & fermion mass $m_f$ [$\times (|q| / 1~\textrm{m})^{-1/2}~\textrm{GeV}$]\\
        \hline
        0 & 40 \\
        10 & 80 \\
        20 & 120 \\
        100 & 300 \\
        1\,000 & 1\,000 \\
        \hline
        \hline
    \end{tabular*}
\end{table}

We see in Fig.~\ref{fig:mrq_compactness} that the compactness of a compact object approaches the range allowed for \ac{ks} black holes, when its fermion mass $m_f$ is larger than the $y$-dependent value given in Table~\ref{tab:compactness} and its central density is given by $\rho_c \gtrsim 10^{23}\times (|q|/{1~\mathrm{m}})^{-2}~\mathrm{g}/\mathrm{cm}^3$. 
In comparison with fermionic compact objects in \ac{gr}, the solutions in \ac{hl} gravity exhibit a systematically higher upper bound on the compactness, indicating that stable equilibrium configurations with larger compactness can be supported in higher $m_f$ and $\rho_c$.
This behavior suggests that considering an exotic particle of higher $m_f$ and $\rho_c$ or considering a large $q$ in the \ac{hl} gravity allows us to expect equilibrium structures which can vanish the compactness gap in \ac{gr}.

\subsection{Compact dark matter candidate in HL gravity}
\label{sec:dm}

A smaller $q$ leads to a smaller minimal \ac{bh} and the compactness of fermionic compact objects is bounded from above $M/R < 1$. Hence, in this section, we focus on smaller $q \leq 1~\mathrm{m}$ and examine the mass of corresponding compact object.

For the case of $q \sim 1~\textrm{m}$, the radius and mass of the minimal \ac{bh} are $r_+^\text{min} = 1~\textrm{m}$ 
and $M_\text{BH}^\text{min} \sim 10^{-4}~\msun$, 
respectively, which corresponds to 
the cold \ac{dm} candidate~\cite{Narain:2006kx}. 
Although this mass of primordial \ac{bh} has been excluded as constituting the majority of cold \ac{dm}, it may still 
survives as a subdominant component of \ac{dm} (see Ref.~\cite{Oncins:2022ydg} and references therein). For the case of $q \sim 1~\textrm{nm}$, on the other hand, 
it alone may account for the entire \ac{dm} abundance, since the mass becomes $M_\text{BH}^\text{min} \sim 10^{-13}~\msun$.

Meanwhile, regarding the Hawking temperature $T_H = (4 \pi r_+)^{-1} (r_+^2 - q^2) / (r_+^2 + 2 q^2)$~\cite{Liu:2011zzu}, we expect that the minimal \ac{bh} does not evaporate, namely, $T_H=0$ because $r_+ = |q|$. Therefore, there may be a population of (near-)minimal \acp{bh} in the Universe as remnants of the evaporation of primordial \acp{bh}. Moreover, no radiation of minimal \ac{bh} and faint radiation of near-minimal \ac{bh} push the lower bound of the mass of primordial \ac{bh} even lower. This implies that much smaller $q$ (i.e. much smaller minimal \ac{bh}), is available as a cold \ac{dm} candidate, which would be challenging to detect with current observational techniques such as microlensing.

In the same manner, a fermionic compact object can have similar mass and radius to those of the minimal \ac{bh} when $m_f$ is high enough and $\rho_c$ is close to $10^{24} \times (|q|/1\ \textrm{m})^{-2}\ \textrm{g/cm}^{3}$. Then, if $q$ is small enough, it is naturally expected that tiny fermionic compact objects of $M \sim M^\mathrm{min}_\mathrm{BH} \ll 1~\msun$ can be formed.
In addition, a compact object of compactness $\sim 1$ may play the role of a cold \ac{dm} candidate, since it does not evaporate and its small size allow it to survive without fatal interactions. Thus, both the (near-)minimal \acp{bh} and tiny fermionic compact objects may contribute to \ac{dm}.

\section{Discussions}
\label{sec:discussion}

Consequently, both the (near-)minimal \acp{bh} and the compact objects of compactness $\sim 1$ are capable of playing the role of \ac{dm}. Moreover, for $q \sim 1\ \textrm{nm}$, the mass of the minimal \ac{bh} becomes $M_\mathrm{BH}^\mathrm{min} \sim 10^{-13}\ \msun$ and the compact object at this scale remains essentially unconstrained by current observation. However, at this small scale of $q$, the fermion mass scale in Table~\ref{tab:compactness} increases dramatically to $\sim 120\ \mathrm{TeV} - 3\ \mathrm{PeV}$, far beyond the energy reach of any existing particle physics experiments.

It is noteworthy that the mass-radius curve shown in Fig.~\ref{fig:mr_mn_ap4} converges toward the extreme point of $M=M_{c}$. This behavior is one of the key motivations of this work and it can be regarded as a distinctive feature of the \ac{ks} solution in the deformed \ac{hl} which exhibits a charged \ac{bh}-like structure with two horizons. As shown in Figs.~\ref{fig:mrq} and \ref{fig:mrq_compactness}, both with and without interaction, a fermionic compact object with smaller fermion mass is similar to one in \ac{gr}, whereas one with larger fermion mass is similar to the minimal \ac{bh} in mass and radius. This is a drastically different feature compared to the GR case.

\begin{figure}
    \centering
    \includegraphics[width=\columnwidth]{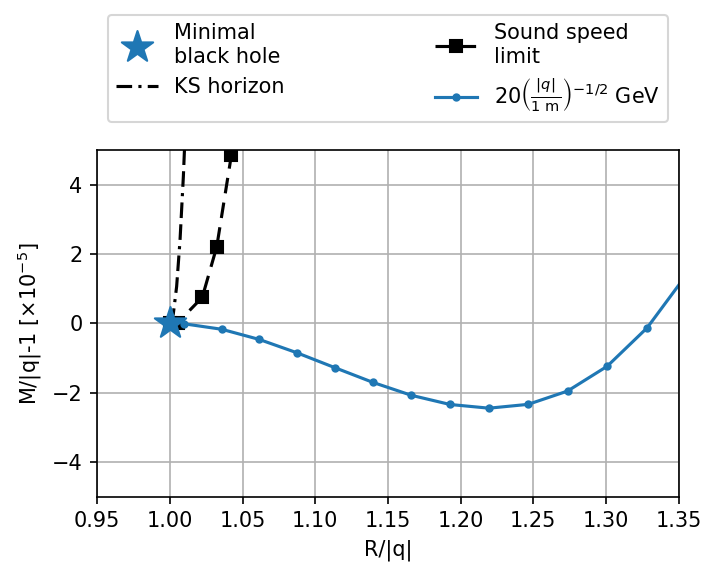}
    \caption{Mass-radius profile of fermionic compact objects with fermion mass $m_f = 20 (|q|/1\ \textrm{m})^{-1/2}\ \textrm{GeV}$ near the minimal \ac{bh} suggests a possible new state of compact object that is more compact than a \acs{ns}. Sound speed limit is also depicted to show that this new compact objects does not exceed the sound speed limit.}
    \label{fig:nearEBH}
\end{figure}

Another intriguing finding of this work is a possible existence of a new class of compact objects in \ac{hl} gravity. We find that the compactness gap may be filled by fermionic compact objects with a fermion mass less than the given value in Table~\ref{tab:compactness}. Fig.~\ref{fig:nearEBH} shows the possibility of an additional branch of compact object which is smaller than \ac{ns} both in mass and radius. Fixing $q \sim 1\ \mathrm{m}$ for simplicity, one finds that the mass and radius of the most massive compact object with fermion mass $20\ \mathrm{GeV}$ without interaction ($y=0$) are $\sim 10^{-3}\ \msun$ and $\sim 20\ \mathrm{m}$, respectively. 
For smaller radius, an additional peak appears in the mass-radius curve that is similar to the minimal \ac{bh} in mass and radius. However, we do not observe this new class of compact object in the cases of $y>0$, which may be attributed to the relatively coarse parameter set to reduce the computational cost. Nevertheless, there still remains the possibility that the interaction term could eliminate this new branch of compact objects. Further detailed investigation will be necessary to clarify this issue.

\acknowledgments 
The authors would like to thank Y.-M.\ Kim, M.-I.\ Park, K.\-Y.\ Kim, and J.\-W.\ Lee for the helpful discussion.
This work is supported by the National Research Foundation of Korea (NRF) grants funded by the Ministry of Science and ICT (MSIT) of the Korea Government (NRF-2021R1A2C1093059, NRF-2020R1C1C1005863, RS-2025-00555178). The work of K.K.\ is partially supported by the Korea Astronomy and Space Science Institute under the R\&D program (Project No.~2026-1-810-00) supervised by the MSIT of the Korea Government. The work of E.J.S.\ and J.J.O.\ is partially supported by the National Institute for Mathematical Sciences under the Primary Research Program (Project No.~B26720000) supported by the MSIT of the Korea Government.

\bibliographystyle{apsrev4-2}
\bibliography{references}

@article{Horava:2008ih,
      author         = "Ho\v{r}ava, Petr",
      title          = "{Membranes at Quantum Criticality}",
      journal        = "JHEP",
      volume         = "2009",
      number         = "03",
      pages          = "020",
      year           = "2009",
      doi            = "10.1088/1126-6708/2009/03/020",
      SLACcitation   = "%%CITATION = ARXIV:0812.4287;%%"
}

@article{Horava:2009uw,
      author         = "Ho\v{r}ava, Petr",
      title          = "{Quantum Gravity at a Lifshitz Point}",
      journal        = "Phys. Rev. D",
      volume         = "79",
      year           = "2009",
      pages          = "084008",
      doi            = "10.1103/PhysRevD.79.084008",
      SLACcitation   = "%%CITATION = ARXIV:0901.3775;%%"
}

@article{Horava:2009if,
      author         = "Ho\v{r}ava, Petr",
      title          = "{Spectral Dimension of the Universe in Quantum Gravity at
                        a Lifshitz Point}",
      journal        = "Phys. Rev. Lett.",
      volume         = "102",
      year           = "2009",
      pages          = "161301",
      doi            = "10.1103/PhysRevLett.102.161301",
      SLACcitation   = "%%CITATION = ARXIV:0902.3657;%%"
}

@article{Deser:1981wh,
      author         = "Deser, Stanley and Jackiw, R. and Templeton, S.",
      title          = "{Topologically Massive Gauge Theories}",
      journal        = "Ann. Phys. (N.Y.)",
      volume         = "140",
      year           = "1982",
      pages          = "372-411",
      doi            = "10.1006/aphy.2000.6013, 10.1016/0003-4916(82)90164-6",
      reportNumber   = "MIT-CTP-964",
      SLACcitation   = "%%CITATION = APNYA,140,372;%%"
}

@article{Deser:1982vy,
      author         = "Deser, Stanley and Jackiw, R. and Templeton, S.",
      title          = "{Three-Dimensional Massive Gauge Theories}",
      journal        = "Phys. Rev. Lett.",
      volume         = "48",
      year           = "1982",
      pages          = "975-978",
      doi            = "10.1103/PhysRevLett.48.975",
      SLACcitation   = "%%CITATION = PRLTA,48,975;%%"
}

@article{Wang:2017brl,
      author         = "Wang, Anzhong",
      title          = "{Ho\v{r}ava gravity at a Lifshitz point: A progress report}",
      journal        = "Int. J. Mod. Phys. D",
      volume         = "26",
      year           = "2017",
      number         = "07",
      pages          = "1730014",
      doi            = "10.1142/S0218271817300142",
      SLACcitation   = "%%CITATION = ARXIV:1701.06087;%%"
}

@article{Mukohyama:2010xz,
      author         = "Mukohyama, Shinji",
      title          = "{Ho\v{r}ava-Lifshitz Cosmology: A Review}",
      journal        = "Class. Quant. Grav.",
      volume         = "27",
      year           = "2010",
      pages          = "223101",
      doi            = "10.1088/0264-9381/27/22/223101",
      reportNumber   = "IPMU10-0120",
      SLACcitation   = "%%CITATION = ARXIV:1007.5199;%%"
}

@article{Kehagias:2009is,
      author         = "Kehagias, Alex and Sfetsos, Konstadinos",
      title          = "{The Black hole and FRW geometries of non-relativistic
                        gravity}",
      journal        = "Phys. Lett. B",
      volume         = "678",
      year           = "2009",
      pages          = "123-126",
      doi            = "10.1016/j.physletb.2009.06.019",
      SLACcitation   = "%%CITATION = ARXIV:0905.0477;%%"
}

@article{Gurtug:2017kqf,
      author         = "Gurtug, O. and Mangut, M.",
      title          = "{Quantum probe of Ho\v{r}ava-Lifshitz gravity}",
      journal        = "J. Math. Phys.",
      volume         = "59",
      year           = "2018",
      number         = "4",
      pages          = "042503",
      doi            = "10.1063/1.5001114",
      SLACcitation   = "%%CITATION = ARXIV:1702.08220;%%"
}

@article{Myung_2010,
   title={ADM mass and quasilocal energy of black hole in the deformed Hořava–Lifshitz gravity},
   volume={685},
   ISSN={0370-2693},
   url={http://dx.doi.org/10.1016/j.physletb.2010.01.073},
   DOI={10.1016/j.physletb.2010.01.073},
   number={4-5},
   journal={Phys. Lett. B},
   publisher={Elsevier BV},
   author={Myung, Y. S.},
   year={2010},
   month={Mar},
   pages={318}
}

@article{Park:2009zra,
      author         = "Park, Mu-In",
      title          = "{The Black Hole and Cosmological Solutions in IR modified
                        Ho\v{r}ava Gravity}",
      journal        = "JHEP",
      volume         = "2009",
      number         = "09",
      year           = "2009",
      pages          = "123",
      doi            = "10.1088/1126-6708/2009/09/123",
      SLACcitation   = "%%CITATION = ARXIV:0905.4480;%%"
}

@article{Son:2010bs,
      author         = "Son, Edwin J. and Kim, Wontae",
      title          = "{Traversable wormhole in the deformed Ho\v{r}ava-Lifshitz
                        gravity}",
      journal        = "Phys. Rev. D",
      volume         = "83",
      year           = "2011",
      pages          = "124012",
      doi            = "10.1103/PhysRevD.83.124012",
      SLACcitation   = "%%CITATION = ARXIV:1011.3596;%%"
}

@article{Kim:2008vi,
    author = "Kim, Wontae and Son, Edwin J. and Yoon, Myungseok",
    title = "{Thermodynamic similarity between the noncommutative Schwarzschild black hole and the Reissner-Nordstrom black hole}",
    doi = "10.1088/1126-6708/2008/04/042",
    journal = "JHEP.",
    volume = "2008",
    number = "04",
    pages = "042",
    year = "2008"
}

@article{Kimetal2021,
  title = {Neutron star structure in Ho\ifmmode \check{r}\else \v{r}\fi{}ava-Lifshitz gravity},
  author = {Kim, Kyungmin and Oh, John J. and Park, Chan and Son, Edwin J.},
  journal = {Phys. Rev. D},
  volume = {103},
  issue = {4},
  pages = {044052},
  numpages = {8},
  year = {2021},
  month = {Feb},
  publisher = {American Physical Society},
  doi = {10.1103/PhysRevD.103.044052},
  url = {https://link.aps.org/doi/10.1103/PhysRevD.103.044052}
}

@misc{Sonlimits,
      title={Limits on the mass of compact objects in Ho\v{r}ava-Lifshitz gravity}, 
      author={Edwin J. Son},
      year={2026},
      eprint={2601.03644},
      archivePrefix={arXiv},
      primaryClass={gr-qc},
      url={https://arxiv.org/abs/2601.03644}, 
}

@article{Liu:2011zzu,
    author = "Liu, Molin and Lu, Junwang",
    title = "{Logarithmic entropy of Kehagias-Sfetsos black hole with self-gravitation in asymptotically flat IR modified Horava gravity}",
    doi = "10.1016/j.physletb.2011.04.006",
    journal = "Phys. Lett. B",
    volume = "699",
    pages = "296--300",
    year = "2011"
}

@article{PhysRevLett.103.091301,
  title = {Solutions to Ho\ifmmode \check{r}\else \v{r}\fi{}ava Gravity},
  author = {L\"u, H. and Mei, Jianwei and Pope, C. N.},
  journal = {Phys. Rev. Lett.},
  volume = {103},
  issue = {9},
  pages = {091301},
  numpages = {4},
  year = {2009},
  month = {Aug},
  publisher = {American Physical Society},
  doi = {10.1103/PhysRevLett.103.091301},
  url = {https://link.aps.org/doi/10.1103/PhysRevLett.103.091301}
}

@article{Bellorin:2014qca,
      author         = "Bellorin, Jorge and Restuccia, Alvaro and Sotomayor,
                        Adrian",
      title          = "{Wormholes and naked singularities in the complete
                        Ho\v{r}ava theory}",
      journal        = "Phys. Rev. D",
      volume         = "90",
      year           = "2014",
      number         = "4",
      pages          = "044009",
      doi            = "10.1103/PhysRevD.90.044009",
      SLACcitation   = "%%CITATION = ARXIV:1404.2884;%%"
}

@article{Tolman:1939jz,
      author         = "Tolman, Richard C.",
      title          = "{Static solutions of Einstein's field equations for
                        spheres of fluid}",
      journal        = "Phys. Rev.",
      volume         = "55",
      year           = "1939",
      pages          = "364-373",
      doi            = "10.1103/PhysRev.55.364",
      SLACcitation   = "%%CITATION = PHRVA,55,364;%%"
}

@article{Oppenheimer:1939ne,
      author         = "Oppenheimer, J. R. and Volkoff, G. M.",
      title          = "{On Massive neutron cores}",
      journal        = "Phys. Rev.",
      volume         = "55",
      year           = "1939",
      pages          = "374-381",
      doi            = "10.1103/PhysRev.55.374",
      SLACcitation   = "%%CITATION = PHRVA,55,374;%%"
}

@article{APR:1998,
      author         = "Akmal, A. and Pandharipande, V. R. and Ravenhall, D. G.",
      title          = "{The Equation of state of nucleon matter and neutron star
                        structure}",
      journal        = "Phys. Rev. C",
      volume         = "58",
      year           = "1998",
      pages          = "1804-1828",
      doi            = "10.1103/PhysRevC.58.1804",
      SLACcitation   = "%%CITATION = NUCL-TH/9804027;%%"
}

@article{Rhoades:1974fn,
    author = "Rhoades, Jr., Clifford E. and Ruffini, Remo",
    title = "{Maximum mass of a neutron star}",
    doi = "10.1103/PhysRevLett.32.324",
    journal = "Phys. Rev. Lett.",
    volume = "32",
    pages = "324--327",
    year = "1974"
}

@article{Kalogera:1996ci,
      author         = "Kalogera, Vassiliki and Baym, Gordon",
      title          = "{The maximum mass of a neutron star}",
      journal        = "Astrophys. J.",
      volume         = "470",
      year           = "1996",
      pages          = "L61-L64",
      doi            = "10.1086/310296",
      SLACcitation   = "%%CITATION = ASTRO-PH/9608059;%%"
}

@article{Margalit:2017dij,
      author         = "Margalit, Ben and Metzger, Brian D.",
      title          = "{Constraining the Maximum Mass of Neutron Stars From
                        Multi-Messenger Observations of GW170817}",
      journal        = "Astrophys. J.",
      volume         = "850",
      year           = "2017",
      number         = "2",
      pages          = "L19",
      doi            = "10.3847/2041-8213/aa991c",
      SLACcitation   = "%%CITATION = ARXIV:1710.05938;%%"
}

@article{Dadhich:2019jyf,
    author = "Dadhich, Naresh",
    title = "{Buchdahl compactness limit and gravitational field energy}",
    doi = "10.1088/1475-7516/2020/04/035",
    journal = "JCAP",
    volume = "2020",
    number = "04",
    pages = "035",
    year = "2020"
}

@article{Giuliani:2007zza,
    author = "Giuliani, Alessandro and Rothman, Tony",
    title = "{Absolute stability limit for relativistic charged spheres}",
    doi = "10.1007/s10714-007-0539-7",
    journal = "Gen. Rel. Grav.",
    volume = "40",
    pages = "1427--1447",
    year = "2008"
}

@ARTICLE{2020SciPy-NMeth,
  author  = {Virtanen, Pauli and Gommers, Ralf and Oliphant, Travis E. and
            Haberland, Matt and Reddy, Tyler and Cournapeau, David and
            Burovski, Evgeni and Peterson, Pearu and Weckesser, Warren and
            Bright, Jonathan and {van der Walt}, St{\'e}fan J. and
            Brett, Matthew and Wilson, Joshua and Millman, K. Jarrod and
            Mayorov, Nikolay and Nelson, Andrew R. J. and Jones, Eric and
            Kern, Robert and Larson, Eric and Carey, C J and
            Polat, {\.I}lhan and Feng, Yu and Moore, Eric W. and
            {VanderPlas}, Jake and Laxalde, Denis and Perktold, Josef and
            Cimrman, Robert and Henriksen, Ian and Quintero, E. A. and
            Harris, Charles R. and Archibald, Anne M. and
            Ribeiro, Ant{\^o}nio H. and Pedregosa, Fabian and
            {van Mulbregt}, Paul and {SciPy 1.0 Contributors}},
  title   = {{{SciPy} 1.0: Fundamental Algorithms for Scientific
            Computing in Python}},
  journal = {Nat. Methods},
  year    = {2020},
  volume  = {17},
  pages   = {261--272},
  adsurl  = {https://rdcu.be/b08Wh},
  doi     = {10.1038/s41592-019-0686-2},
}

@Inbook{Hairer1993,
author="Hairer, Ernst and Wanner, Gerhard and N{\o}rsett, Syvert P.",
title="Runge-Kutta and Extrapolation Methods",
bookTitle="Solving Ordinary Differential Equations I: Nonstiff Problems",
year="1993",
publisher="Springer Berlin Heidelberg",
address="Berlin, Heidelberg",
pages="129--353",
isbn="978-3-540-78862-1",
doi="10.1007/978-3-540-78862-1_2",
url="https://doi.org/10.1007/978-3-540-78862-1_2"
}

@article{LIGOScientific:2024elc,
    author = "Abac, A. G. and Abbott, R. and Abouelfettouh, I. and others",
    collaboration = "LIGO Scientific, KAGRA, VIRGO",
    title = "{Observation of Gravitational Waves from the Coalescence of a 2.5{\textendash}4.5 M $_{⊙}$ Compact Object and a Neutron Star}",
    reportNumber = "LIGO-P2300352",
    doi = "10.3847/2041-8213/ad5beb",
    journal = "Astrophys. J. Lett.",
    volume = "970",
    number = "2",
    pages = "L34",
    year = "2024"
}

@article{Ozel_2010,
doi = {10.1088/0004-637X/725/2/1918},
url = {https://doi.org/10.1088/0004-637X/725/2/1918},
year = {2010},
month = {dec},
publisher = {The American Astronomical Society},
volume = {725},
number = {2},
pages = {1918},
author = {{\"O}zel, Feryal and Psaltis, Dimitrios and Narayan, Ramesh and McClintock, Jeffrey E.},
title = {THE BLACK HOLE MASS DISTRIBUTION IN THE GALAXY},
journal = {Astrophys. J.}
}

@article{Fishbach_2020,
doi = {10.3847/2041-8213/aba7b6},
url = {https://doi.org/10.3847/2041-8213/aba7b6},
year = {2020},
month = {aug},
publisher = {The American Astronomical Society},
volume = {899},
number = {1},
pages = {L8},
author = {Fishbach, Maya and Essick, Reed and Holz, Daniel E.},
title = {Does Matter Matter? Using the Mass Distribution to Distinguish Neutron Stars and Black Holes},
journal = {Astrophys. J. Lett.}
}

@article{Fishbach_2024,
author = {Maya Fishbach },
title = {Mystery in the "mass gap"},
journal = {Science},
volume = {383},
number = {6680},
pages = {259-260},
year = {2024},
doi = {10.1126/science.adn1869},
}

@article{Fishbach2017,
  author  = {Fishbach, Maya and Holz, Daniel E. and Farr, Will M.},
  title   = {Does the Black Hole Mass Distribution Support a Hierarchical Merger Scenario?},
  journal = {Astrophys. J. Lett.},
  volume  = {840},
  pages   = {L24},
  year    = {2017},
  doi     = {10.3847/2041-8213/aa7045}
}

@article{GerosaBerti2017,
  author  = {Gerosa, Davide and Berti, Emanuele},
  title   = {Are merging black holes born from stellar collapse or previous mergers?},
  journal = {Phys. Rev. D},
  volume  = {95},
  pages   = {124046},
  year    = {2017},
  doi     = {10.1103/PhysRevD.95.124046}
}

@article{Abbott2020Population,
  author  = {Abbott, R. and others},
  collaboration = {LIGO Scientific Collaboration and Virgo Collaboration},
  title   = {Population Properties of Compact Objects from the Second LIGO--Virgo Gravitational-Wave Transient Catalog},
  journal = {Astrophys. J. Lett.},
  volume  = {913},
  pages   = {L7},
  year    = {2020},
  doi     = {10.3847/2041-8213/abc7fc}
}

@article{Fishbach2024,
  author  = {Fishbach, Maya and others},
  title   = {The Black Hole--Neutron Star Mass Gap},
  journal = {Living Rev. Relativ.},
  volume  = {27},
  pages   = {2},
  year    = {2024},
  doi     = {10.1007/s41114-024-00068-9}
}

@article{Fishbach2020,
  author  = {Fishbach, Maya and Farr, Will M. and Holz, Daniel E.},
  title   = {Does the Black Hole Mass Gap Exist?},
  journal = {Astrophys. J. Lett.},
  volume  = {891},
  pages   = {L31},
  year    = {2020},
  doi     = {10.3847/2041-8213/ab7247}
}

@misc{Oncins:2022ydg,
      title={Constraints on PBH as dark matter from observations: a review}, 
      author={Marc Oncins},
      year={2022},
      eprint={2205.14722},
      archivePrefix={arXiv},
      primaryClass={astro-ph.CO},
      url={https://arxiv.org/abs/2205.14722}, 
}

@article{Narain:2006kx,
    author = "Narain, Gaurav and Schaffner-Bielich, Jurgen and Mishustin, Igor N.",
    title = "{Compact stars made of fermionic dark matter}",
    doi = "10.1103/PhysRevD.74.063003",
    journal = "Phys. Rev. D",
    volume = "74",
    pages = "063003",
    year = "2006"
}

@article{Mariani:2023,
    author = {Mariani, Mauro and Albertus, Conrado and Alessandroni, M del Rosario and Orsaria, Milva G and Pérez-García, M Ángeles and Ranea-Sandoval, Ignacio F},
    title = {Constraining self-interacting fermionic dark matter in admixed neutron stars using multimessenger astronomy},
    journal = {Mon. Not. R. Astron. Soc.},
    volume = {527},
    number = {3},
    pages = {6795-6806},
    year = {2023},
    month = {11},
    issn = {0035-8711},
    doi = {10.1093/mnras/stad3658}
}

@article{Valdez-Alvarado:2013,
  title = {Dynamical evolution of fermion-boson stars},
  author = {Valdez-Alvarado, Susana and Palenzuela, Carlos and Alic, Daniela and Ure\~na-L\'opez, L. Arturo},
  journal = {Phys. Rev. D},
  volume = {87},
  issue = {8},
  pages = {084040},
  numpages = {12},
  year = {2013},
  month = {Apr},
  publisher = {American Physical Society},
  doi = {10.1103/PhysRevD.87.084040},
  url = {https://link.aps.org/doi/10.1103/PhysRevD.87.084040}
}

@article{Mukhopadhyay:2016,
  title = {Quark stars admixed with dark matter},
  author = {Mukhopadhyay, Payel and Schaffner-Bielich, J\"urgen},
  journal = {Phys. Rev. D},
  volume = {93},
  issue = {8},
  pages = {083009},
  numpages = {10},
  year = {2016},
  month = {Apr},
  publisher = {American Physical Society},
  doi = {10.1103/PhysRevD.93.083009},
  url = {https://link.aps.org/doi/10.1103/PhysRevD.93.083009}
}

\end{document}